\documentclass[conference]{IEEEtran}
\IEEEoverridecommandlockouts

\usepackage{multirow}
\usepackage[flushleft]{threeparttable}
\usepackage{latexsym}
\usepackage{amsmath}
\usepackage{amssymb}
\usepackage{amsthm}
\usepackage{graphicx}
\usepackage{epsfig}
\usepackage{amssymb}
\usepackage{subcaption}
\usepackage{mwe}
\usepackage{float}
\usepackage{mathtools}
\usepackage{algorithm}
\usepackage[noend]{algpseudocode}
\usepackage[utf8]{inputenc}
\usepackage[english]{babel}
\usepackage{fancyhdr}
\usepackage{mleftright}
\usepackage{courier}
\usepackage{array}
\usepackage{booktabs}
\usepackage{syntonly}
\usepackage{xcolor}
\usepackage{colortbl}
\usepackage{lipsum}
\usepackage{epstopdf}
\usepackage{environ}
\usepackage{relsize}
\usepackage{multirow}
\usepackage{diagbox}
\usepackage{color} 
\usepackage[normalem]{ulem} 
\usepackage[makeroom]{cancel} 
\usepackage{bm}
\usepackage{enumitem}
\usepackage[normalem]{ulem}
\PassOptionsToPackage{hyphens}{url}\usepackage{hyperref}
\def\BibTeX{{\rm B\kern-.05em{\sc i\kern-.025em b}\kern-.08em
    T\kern-.1667em\lower.7ex\hbox{E}\kern-.125emX}}

\makeatletter
\let\OldStatex\Statex
\renewcommand{\Statex}[1][3]{%
  \setlength\@tempdima{\algorithmicindent}%
  \OldStatex\hskip\dimexpr#1\@tempdima\relax}
\makeatother

\NewEnviron{myequation}[1][]{%
\begin{equation}
\scalebox{#1}{$\begin{aligned}\BODY\end{aligned}$}
\end{equation}
}
\NewEnviron{mymultiline}[1][]{%
\scalebox{#1}{\begin{align}
\BODY
\end{align}}
}

\newcommand{\mb}[1]{{\mathbf{#1}}}
\newcommand{\m}[1]{\mathrm{#1}}

\newcommand{\mc}[1]{{\mathcal{#1}}}

\newcommand{\PreserveBackslash}[1]{\let\temp=\\#1\let\\=\temp}
\newcolumntype{C}[1]{>{\PreserveBackslash\centering}p{#1}}
\newcolumntype{R}[1]{>{\PreserveBackslash\raggedleft}p{#1}}
\newcolumntype{L}[1]{>{\PreserveBackslash\raggedright}p{#1}}

\begin{document}

\title{A Reinforcement Learning-based Volt-VAR Control Dataset and Testing Environment}



\author{\IEEEauthorblockN{1\textsuperscript{st} Yuanqi Gao}
\IEEEauthorblockA{\textit{University of California, Riverside} \\
Riverside, CA, USA \\
ygao024@ucr.edu}
\and
\IEEEauthorblockN{2\textsuperscript{nd} Nanpeng Yu}
\IEEEauthorblockA{\textit{University of California, Riverside} \\
Riverside, CA, USA \\
nyu@ece.ucr.edu}
}

\maketitle

\begin{abstract}
To facilitate the development of reinforcement learning (RL) based power distribution system Volt-VAR control (VVC), this paper introduces a suite of open-source datasets for RL-based VVC algorithm research that is sample efficient, safe, and robust. The dataset consists of two components: 1. a Gym-like VVC testing environment for the IEEE-13, 123, and 8500-bus test feeders and 2. a historical operational dataset for each of the feeders. Potential users of the dataset and testing environment could first train an sample-efficient off-line (batch) RL algorithm on the historical dataset and then evaluate the performance of the trained RL agent on the testing environments. This dataset serves as a useful testbed to conduct RL-based VVC research mimicking the real-world operational challenges faced by electric utilities. Meanwhile, it allows researchers to conduct fair performance comparisons between different algorithms.
\end{abstract}


\IEEEpeerreviewmaketitle

\section{Introduction}
Volt-VAR control (VVC) is a key component of the advanced distribution system management system. As the power distribution systems are transitioning into the era of high distributed generation penetration, the legacy VVC technologies need to be improved to accommodate these changes and continue to provide high-quality electric power to the end-use customers.

In recent years, data-driven VVC approaches have seen a tremendous developments due to their capabilities to learn from operational data and superior computation efficiency. Among the proposed methods, reinforcement learning (RL) have been studied in great depth as a promising solution to the VVC problem. RL is a class of machine learning algorithms that learns optimal control policies in Markov decision processes (MDPs) \cite{sutton2018reinforcement} and is suited for complex sequential decision-making problems. In the VVC literature, tabular Q-learning \cite{Xu2012Multiagent}, batch RL with training data augmentation \cite{xu2019optimal}, safe RL in constrained MDP \cite{Wei2019Safe}, multi-agent deep Q learning \cite{Zhang2020deep}, multi-agent consensus RL \cite{gao2021consensus} have been proposed.

Compared with physical model-based approaches, RL does not rely on accurate and reliable distribution system model to compute control actions but learn from data. Furthermore, the computation time of RL algorithms in real-time execution is much shorter than model-based controllers. Despite these advantages, there still exists large gap between the theory and practice of RL-based VVC. First, to learn a good policy, many RL algorithms need a substantial amount of training data, which are not always available. Second, building accurate and reliable training environment is not feasible for most of the electric utilities. Third, many RL algorithms cannot guarantee that the voltage or current magnitudes are always within the acceptable range. This is because deep RL algorithms parameterize policies by function approximators which may produce unsafe actions in some states. Some proposed RL-based VVC techniques have addressed the aforementioned problems. However, due to the lack of standardized test dataset, the results are usually difficult to replicate. Furthermore, it is often challenging to fairly compare the performance between existing and newly proposed algorithms.

Motivated by these practical challenges, we develop an open-source dataset for developing and testing RL-based VVC algorithms. This dataset adopts the IEEE-13, 123, and 8500-bus test feeders \cite{Kersting2001testfeeders} \cite{schneider2017analytic}, which was implemented by OpenDSS and wrapped as Gym-like VVC testing environments. To facilitate the development of sample-efficient offline RL, we also generated a set of historical operational data for each of the test feeders. These data should be used for offline training. Once an RL algorithm is trained, the VVC testing environment should be treated as a real world distribution system to be controlled. Therefore, it is intended for quantifying online performance metrics such as safety and optimality during \textit{both} exploration and the final algorithm convergence. We hope the dataset can serve as a useful benchmark for the research and development community. The code of this paper can be accessed at \cite{code}, where the power flow program is based on \cite{dsscircuits}.

The rest of this paper is organized as follows. Section II provides an overview of RL-based VVC methods. Section III describes the proposed datasets. Section IV provides preliminary benchmark results. Section V concludes this paper.

\section{An Overview of RL-based VVC}
This section provides an overview of VVC problem formulation and the basics of reinforcement learning. Then we discuss the practical challenges for RL-based VVC.

\subsection{Volt-VAR Control Problem Formulation}
The VVC module adjusts the tap positions of voltage regulators and on-load tap changers (OLTCs), as well as the on/off status of field and substation capacitor banks. The objective of VVC is typically maintaining the desired voltage profile, reducing network loss, or regulating the power factor. VVC can be accomplished by classical methods or advanced control methods \cite{EPRI2011VVCassessment}. Classical methods use local measurements and line-drop compensators to control voltage regulators and OLTCs; the substation capacitors and field capacitors' on-off status are controlled by a set of pre-defined rules according to their local voltage measurements. The classical methods have served the electric utility industry for many years and were effective for legacy distribution systems.

To improve the VVC performance, advanced control algorithms have been studied in the literature. We introduce the VVC problem setup assuming balanced three-phase networks. Nevertheless, our datasets are designed for unbalanced three-phase systems. 
Consider a distribution network with $N$ buses. Bus 1 denotes the substation. At each bus $i$ and time $t$, the nodal voltage magnitude, magnitude of nodal current injection, real and reactive power injection are denoted as $V^i_t$, $I^i_t$, $p^i_t$, and $q^i_t$, respectively.
The real and reactive power flow for line $(i,j)$ connecting bus $i$ and $j$ at time $t$ is denoted as $p^{ij}_t$ and $q^{ij}_t$. The substation voltage regulators and field OLTCs between bus $i$ and $j$ are modeled as follows \cite{gao2021consensus}:
\begin{align}\label{oltc}
    (V^{j}_t/u^{ij}_t)^2 = (V^{i}_t)^2 - 2(r^{\ell}p^{ij}_t + x^{\ell}q^{ij}_t) + (r^{\ell2} + x^{\ell2})I^{ij}_t
\end{align}
where $I^{ij}_t$ is the squared current magnitude of line $(i,j)$; $u^{ij}_t$ is the turns ratio of the OLTC 
The turns ratio changes 5/8\% per tap-change; 33 tap positions divides the 0.9 - 1.1 p.u. voltage control range evenly. Substation and field capacitors are modeled as voltage-dependent reactive power sources. The reactive power output is given by:
\begin{align}
q^{i,\m{cap}}_t = h^i_t \cdot M^{\m{cap}} \cdot (V_t^i)^2,
\end{align}
where $h^i_t \in \{0,1\}$ is the status of capacitor; $M^{\m{cap}}$ is the capacitor reactive power output at rated voltage. 
 
The complete power flow model are given by \eqref{oltc}-\eqref{eq:ref_volt}.
\begin{align}
&p^i_t = \sum_{j:i\rightarrow j} p^{ij}_t - \sum_{j:j\rightarrow i} (p^{ji}_t-r^{\ell}l^{ij}_t) \quad \forall i = 2, ..., N\\
&  q^i_t + q_t^{i,\m{cap}} = \sum_{j:i\rightarrow j} q^{ij}_t - \sum_{j:j\rightarrow i} (q^{ji}_t-x^{\ell}l^{ij}_t) \; \forall i = 2, ..., N\\
&l^{ij}_t=[(p^{ij}_t)^2 + (q^{ij}_t)^2]/(V_t^i)^2 \qquad \forall ij\in \mc{E}\\
&V^1_t = 1 \m{p.u.} + x^{\m{reg}}_t \cdot M^{\m{reg}} 
\label{eq:ref_volt}
\end{align}
The control objective of VVC is to set the voltage regulator/OLTC turns ratio and capacitor bank on-off status to improve voltage profile and reactive power level for each discrete time stamp $t = 1, 2, ...$. The problem can be formulated as:
\begin{equation}\label{vvc_problem}
\begin{aligned}
\underset{\mb{u}, \mb{h}}{\min} &
& & \sum_t \lambda_V f_V(t) + \lambda_{PF} f_{PF}(t) + \lambda_L f_L(t) + \\
& & & \lambda_u |\mb{u}_t - \mb{u}_{t-1}| + \lambda_h |\mb{h}_t - \mb{h}_{t-1}| \\
\text{s.t.} &
& &  \eqref{oltc}-\eqref{eq:ref_volt}
\end{aligned}
\end{equation}
where $\mb{u} = [\mb{u}_1, \mb{u}_2, ... ]$, $\mb{u}_t = [u_t^1, u_t^2, ... ]$ collects the turns ratio at all timestamps and all devices. $f_V$, $f_{PF}$, and $f_L$ are control objectives related to voltage, power factor, and network loss, respectively. The remaining two terms account for the switching cost. The VVC problem is a sequential decision making problem with uncertainties (load, distributed generations, etc.). Next we briefly review reinforcement learning which can be used to solve these problems.

\subsection{Overview of Reinforcement Learning}
Reinforcement learning (RL) \cite{sutton2018reinforcement} solves a class of sequential decision problems known as Markov decision processes (MDPs). An MDP is a tuple ($\mc{S}$, $\mc{A}$, $P$, $r$, $\gamma$) containing a state space $\mc{S}$, an action space $\mc{A}$, a state transition probability function $P(s^\prime|s,a)$, a reward function $r(s,a): \mc{S}\times\mc{A}\mapsto \mathbb{R}$, and a discount factor $\gamma\in (0,1)$. The RL agent interacts with the environment by taking an action $A_t$ at each state $S_t$. the MDP then returns a reward following each action $R_{t+1}=r(S_t,A_t)$ and transitions to another state $S_{t+1}$. The above process continues. The goal of the RL agent is to learn a policy $\pi(a|s)$, which maps states to actions, such that the state value function $v^\pi(s) = \mathbb{E}_{\pi} \left[ \sum_{t=0}^T \gamma^t R_{t+1}|S_0=s \right]$ is maximized. The discount factor $\gamma$ properly lowers future rewards according to the learning objective. $T$ is the length of an episode, which may be infinite. A related function is the action value function defined as $q^\pi(s,a) = \mathbb{E}_{\pi} \left[ \sum_{t=0}^T \gamma^t R_{t+1}|S_0=s, A_0=a \right]$.

The optimal value functions satisfy the Bellman optimality equations:
\begin{align}
q^*(s, a) &= r(s, a) + \gamma \mathbb{E}_{P(s^\prime|s ,a)}[\m{max}_{a^\prime} q^*(s^\prime ,a^\prime)] \\
v^*(s) &= \m{max}_{a} r(s, a) + \gamma \mathbb{E}_{P(s^\prime|s ,a)}[ v^*(s^\prime)]
\end{align}
 
In practice, the functional form of $r(s,a)$ and $P(s'|s,a)$ in an MDP are typically unknown or difficult to specify. RL algorithms must learn from the sample trajectories $S_0, A_0, R_1, S_1, A_1, R_2, ...$ collected by interacting with the environment. Some MDPs have intractably high-dimensional state-action spaces. Deep RL approaches this problem by learning features of the state action along with value functions. 

To apply RL method in VVC problems, we start by identifying the state (e.g. network operating condition), action (e.g. device tap settings), and reward (e.g. control objectives). Then train appropriate RL algorithms either on offline dataset or by interacting with an simulation environment. 
In the next subsection, we summarize the important challenges for applying RL to VVC. Then we discuss a few promising ways to solve these problems which motivates the proposed dataset.

\subsection{Challenges}
When applying RL to the VVC problem, a few practical challenges need to be addressed \cite{gao2021challenges}:
\subsubsection{Sample efficiency}
RL algorithms usually require a large number of agent-environment interactions to learn a good policy. This is difficult for VVC problems since data are limited. Off-policy RL algorithms are capable of re-using previously collected operational experiences. Thus one method to improve the sample efficiency is to train an off-policy RL agent on the historical dataset before interacting with the grid. An open-source benchmark should provide such an offline dataset to support the development of off-policy algorithms.

\subsubsection{Availability of accurate simulation environment}
Another way to circumvent the sample efficiency issue is to build a simulation environment for the VVC problem. Then train RL algorithms on this simulation environment before migrating to the field. Unfortunately, most of the electric utilities do not have accurate model parameters, such as line impedances, for the distribution networks. Worse yet, a majority of the distribution feeders' smart meter coverage is limited. In practice, estimating the distribution network physical model is still a challenging task. 

\subsubsection{Safety}
To operate in distribution systems with many critical loads and infrastructures, RL policy must be safe and robust especially in unforeseen operation conditions. Therefore, the safety and reliability should be an important evaluation criteria for RL-based VVC. The dataset should be designed in a way that allows evaluation of reliability both during early stage of learning and after convergence.

In the next subsection, we describe the proposed dataset that attempt to cover the above practical challenges.

\section{The Proposed Dataset}
This section provides the details of the  open-source dataset. First we present the RL problem formulation. Next we discuss the VVC testing environment and offline datasets.
\subsection{Problem Formulation}
This subsection identifies the state, action, reward, and control horizons of the VVC setup. In the dataset, half-hour time interval is assumed.
\subsubsection{State}
Three state space formulations are provided.

State option 1: $S_t = [p_t, q_t, Tap_{t-1}, t]$, where $p_t = \{p^1_{t-1}, p^i_t| i\in \mc{M} \}$ and similarly for $q_t$; $Tap_{t-1}$ is the list of VVC device status at time $t-1$. This option assumes the AMI data are available at the same frequency as the control devices' operating frequency. This is an ideal case and has been used for some existing RL-based VVC papers. However, AMI data are typically only available by the end of each operating day, due to communication bottleneck. Therefore we provide option 2 and 3 which are more realistic ways of forming the state space.

State option 2: $S_t = [p^1_{t-1}, q^1_{t-1}, Tap_{t-1}, t]$. This formulation only includes SCADA measurements, which can be readily adopted in practice if the system operator can only access SCADA data in real time. However, it does not allow the RL agent to form an estimate for the load condition of different regions in the distribution feeder.

State option 3: $S_t = [p_{t-48}, q_{t-48}, Tap_{t-1}, t]$ where $p_{t-48} = \{p^1_{t-1}, p^i_{t-48}|i\in \mc{M}\}$ and similarly for $q_{t-48}$. The SCADA measurements are fully utilized, whereas the AMI measurements are taken from the same half-hour at the previous day to account for the communication delay.

Some feature engineering of states is provided by default. First, all real and reactive power are normalized by their average value across time. Second, the global time stamp $t$ has been encoded as a list of sin-cos periodic variables: $t\mapsto [\cos(2\pi t / T_i), \sin(2\pi t/T_i)]$, where $T_i$ takes the values of $24\times 2$ and $168 \times 2$ which corresponds to daily and weekly patterns, respectively. Third, to reduce the dimensionality of the 8500-bus test feeder, the AMI data have been averaged for every 10 smart meters. This reduces the state dimension from the order of thousands to hundreds. 

\subsubsection{Reward}
To reflect different control objectives, we provide the following reward formulations:

Reward option 1: $R_{t+1} = -\beta_1 \sum_{i\in \mc{M}} |V^i_t - 1.0 \m{p.u.}| - \beta_2 p^l_t - \beta_3 \sum_{i} |Tap^i_{t-1} - Tap^i_t|$ where $p^l_t$ is the network total line loss in per unit; $\beta_1 = 0.5, \beta_2 = 1.0, \beta_3 = 0.1$ are default reward coefficients. 
In practice, the network total line loss might be difficult to measure. In addition, voltage limit constraints could not be strictly enforced. The following reward options are useful alternatives.

Reward option 2: The same setup as reward option 1. The only difference is that the network loss term is not included. 

Reward option 3: $R_{t+1} = -\beta_1 \sum_{i\in \mc{M}} [\mathbb{I}(V^i_t>1.05 \m{p.u.}) +\mathbb{I}(V^i_t<0.95 \m{p.u.})] - \beta_2 p^l_t - \beta_3 \sum_{i} |Tap^i_{t-1} - Tap^i_t|$ where $\mathbb{I}(condition)$ is the indicator function which equals 1 if the $condition$ equals true and 0 otherwise. Compared with reward option 1, this option only penalizes voltage deviation outside the ANSI C84.1 Service Voltage Range A (i.e. $\pm 5\%$) \cite{short2003electric}.

Reward option 4: The same as reward option 3. The only difference is that the network loss term is not included. 

\subsubsection{Action}
The action is defined as changing the tap positions of voltage regulators, OLTCs, and capacitors to the updated values at the current timestamp.

\subsubsection{Episode}
The VVC problem does not have a natural termination state which separates the whole time into different control episodes. Therefore our default VVC environment is non-episodic.

\subsection{Test feeders}
We adopt the IEEE 13, 123, and 8500-bus test feeders \cite{Kersting2001testfeeders} \cite{schneider2017analytic} in the testing environment. Table. \ref{table_testfeeders} provides the summary statistics of the three test feeders.
\begin{table}[h]
\centering
\begin{threeparttable}
	\caption{Summary of Test Feeders}
	\begin{tabular}{p{1.8cm} C{1.5cm} C{1.5cm} C{1.5cm}}
		\toprule  
		           & 13-bus & 123-bus & 8500-bus \\ \hline
		 \# of loads & 9 & 85 & 1177 \\
		 \# of LTC$^\dagger$ & 1 & 5 & 12 \\
		 \# of capacitors$^\dagger$ & 2 & 4 & 10 \\
		\bottomrule
	\end{tabular}
	\label{table_testfeeders}
	\begin{tablenotes}
      \small
      \item[$\dagger$] We only count the number of independently controllable VVC devices. If a three-phase LTC or three-phase capacitor is gang-operated, we only count it as one device. 
    \end{tablenotes}
\end{threeparttable}
\end{table}

We modify the VVC devices slightly as follows. For the 13-bus test feeder, the three independent voltage regulator controllers are merged as one. Thus there is only one independently controllable LTC. The capacitor control logic is based on the 118/122V voltage setting. That is, in the 120V voltage basis, switch on if the voltage is below 118V; switch off if the voltage is above 122V; remain the same status otherwise. For the 123-bus test feeder, we change regulator 4 to gang-controlled with $R=0.6$ and $X=1.3$. The capacitor control is 122/126V. For the 8500-bus test feeder, the control logic for capacitor 0 and 1 are changed to 120/124V voltage type. The LTC control and other capacitor control are not modified. Capacitor 3 is assumed to be open for all timestamps. However, it can be controlled by the RL algorithm. Finally, we only design the balanced load and constant power factor case for the 8500-bus test feeder.

\subsection{Operational Data}
To train offline RL algorithms, four sets of operational data are generated: (1) load data, (2) LTC tap and capacitor status, (3) voltage data, and (4) SCADA data. 

\subsubsection{Load Data} The load data (kW and kVAr) for each load at each test feeder are generated by multiplying the original single snapshot load of the test feeders with a normalized load time series. The load time series were derived from the London smart meter dataset \cite{londondata}, which contains half-hourly smart meter kWh data for over 5,000 customers during the years 2011-2014. We process the dataset as follows. 
\begin{enumerate}[wide, labelwidth=!, labelindent=0pt]
\item We select 2011-08-01 00:00:00 to 2014-02-28 00:00:00 as our study period as it has relatively small amount of missing values. This produces 27,649 half-hourly timestamps.
\item We select a sufficient amount of customers whose kWh reading has less than 10\% of missing values over the study period. We impute the remaining missing values using the matrix factorization algorithm \cite{fancyimpute}.
\item For each load of the 13 and 123-bus test feeder, we sum the kWh reading over 5 customers at each timestamp.
The summed kWh reading was then normalized by dividing its time-average before multiplying the snapshot load of the test feeders. This preserves the spatial loading pattern of the test feeders while having a realistic loading level. 
\item For each load of the 8500-bus teset feeder, we again multiply the load of the test feeder by the normalized kWh reading as the load time series. The difference is that only 1 customer's data is used for each load rather than 5. 
\end{enumerate}
\subsubsection{LTC Tap, Capacitor Status, Voltage, and SCADA Data}
The LTC tap, capacitor status, voltage, and SCADA data are solved by the OpenDSS static power flow program. The control logic for LTC and capacitors were discussed in the previous subsection. Since the time interval is relatively long (30 minutes), each timestamp is treated as an independent power flow analysis. This is done by setting the control mode of OpenDSS to STATIC. 

Finally, both voltage and load data is rounded to the first decimal place, similar to the real-world AMI data.

\section{Benchmarking Results}
In this section, we provide benchmarking results of two baseline RL algorithms on the test feeders. We also provide the key performance metrics for the baseline algorithms. The code for the results can be found at \cite{code}.

\subsection{Baseline RL Algorithms}
Two baseline deep RL algorithms are implemented: soft actor-critic (SAC) \cite{SAC} and deep Q-learning (DQN) \cite{mnih2015human}. We modify the original algorithmic designs so that they are more suitable for the VVC problem.

\subsubsection{Soft Actor-Critic}
For the SAC algorithm, we use an device-decoupled neural network structure \cite{Wei2019Safe} which scales linearly with the number of devices rather than exponentially. To encode the ordering relationship between different tap positions in neural network-based policies, we adopt the ordinal encoding layer of discrete outputs \cite{DCAOP}. 

\subsubsection{Deep Q Learning}
\begin{figure}[h]
	\centering
	\includegraphics[width=6.5cm]{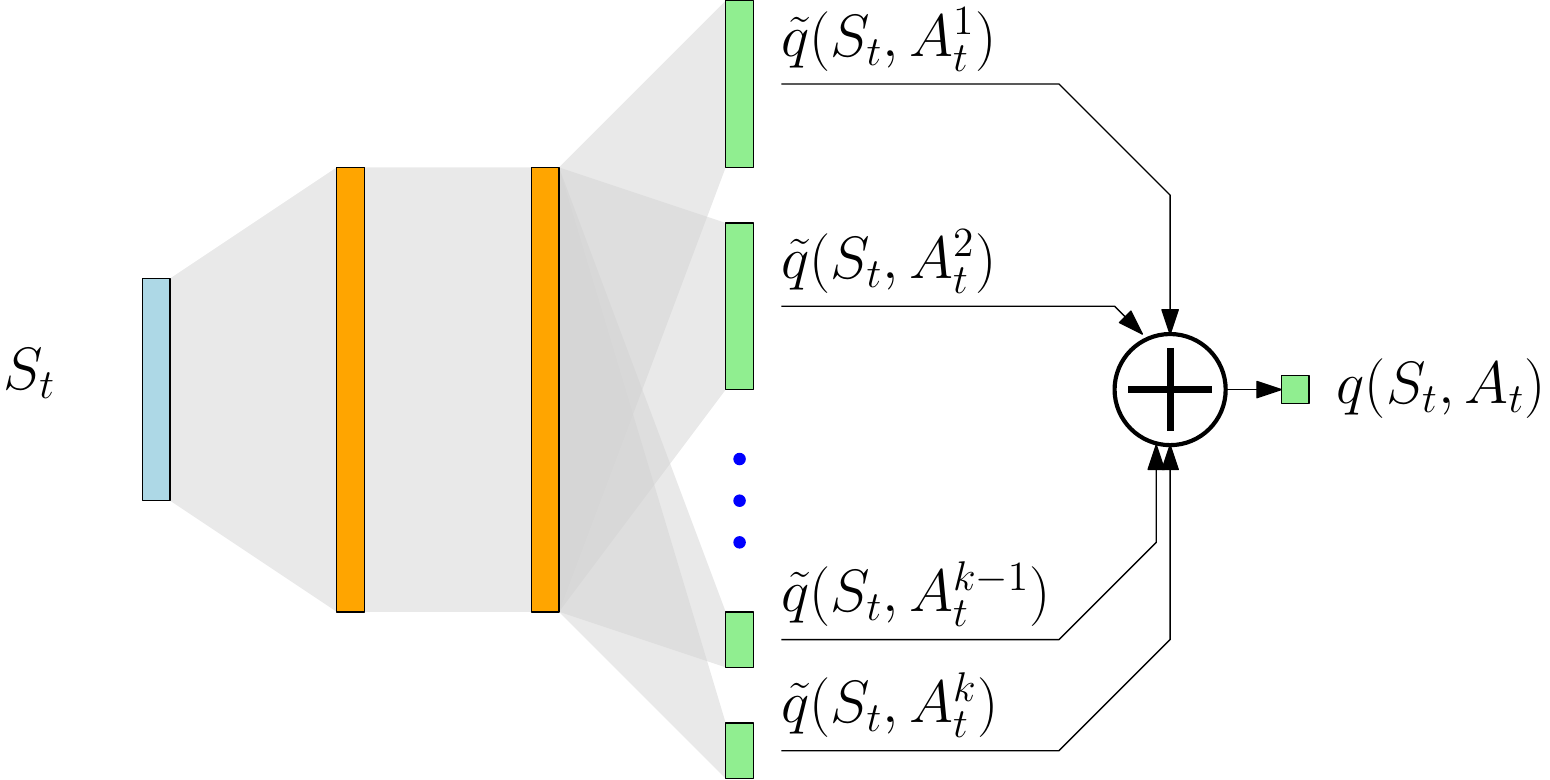}
	\caption{Deep Q network architecture for VVC}
	\label{fig_qnet}
\end{figure}

We modify the DQN algorithm to handle the large action space. First, we introduce the Q network architecture shown in Figure \ref{fig_qnet}.
The hidden layers' activations are sent separately to multiple heads, one for each VVC device. The dimension of each head equals the number of feasible tap settings for that device. We denote the output of head $i$ as $\tilde{q}(S_t, A^i_t)$. Then, we restrict the functional form of the full Q value to be the sum of heads:
\begin{align}
    q(s, a) = \sum_{i=1}^{k} \tilde{q}(s, a^i)
\end{align}
This modified Q network scales linearly with the number of devices $k$. Finally, to efficiently perform the $\max$ and $\arg\max$ operations during training and evaluation, we swap the max and the sum operations:
\begin{align}\label{eq_swap}
    \max_a q(s, a) & = \max_{a} \sum_{i=1}^{k} \tilde{q}(s, a^i) = \sum_{i=1}^{k} \max_{a^i} \tilde{q}(s, a^i)
\end{align}
\eqref{eq_swap} shows that the maximum of Q values can be obtained by maximizing individual heads, hence significantly reducing the computation time. We found this DQN works reasonably well for the test cases. We encourage researchers to develop better algorithms that outperform the baseline algorithms.

\subsection{Benchmarking Results}
\begin{figure}[h]
	\centering
	\includegraphics[width=8.8cm]{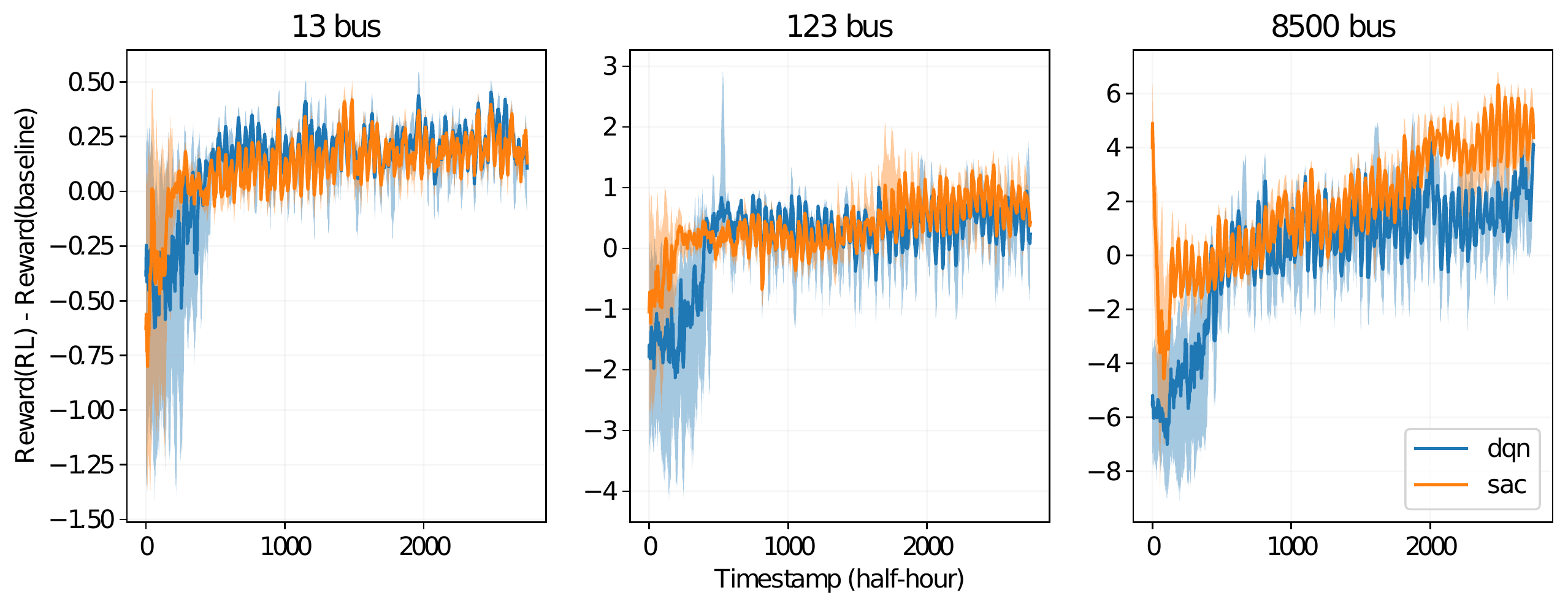}
	\caption{Different between the reward of RL algorithms and that of the default control logic.}
	\label{fig_bus_all}
\end{figure}

Table. \ref{table_params} summarizes the hyperparameters for the test cases, which are used for all three test cases. In particular, the pre-train are performed on the offline experiences before agent-environment interactions. 
\begin{table}[h]
\centering
	\caption{Hyperparameters of Benchmark Algorithms}
	\begin{tabular}{L{3.9cm} L{1.8cm} L{1.8cm}}
		\toprule  
		& SAC & DQN \\ \hline
		hidden layer sizes & (120, 120) & (120, 120)  \\ reward scale & 5.0 & 5.0 \\
		     discount factor ($\gamma$) & 0.95 & 0.95 \\ 
		     batch size & 64 & 64 \\
		     learning rate &  0.0005 & 0.0005 \\ 
		     pre-train steps & 100 & 100 \\
		     temperature parameter ($\alpha$ in \cite{SAC}) & 0.2 & - \\
		     smoothing coefficient ($\tau$ in \cite{SAC}) & 0.01 & - \\
		     copy steps ($C$ in \cite{mnih2015human}) & - & 30 \\
		     epsilon length ($\epsilon$-greedy in \cite{mnih2015human}) & - & 500 \\
		     epsilon max ($\epsilon$-greedy in \cite{mnih2015human})& - & 1.0 \\
		     epsilon min ($\epsilon$-greedy in \cite{mnih2015human})& - & 0.02 \\
		\bottomrule
	\end{tabular}
	\label{table_params}
\end{table}
Figure \ref{fig_bus_all} shows the VVC performance of RL algorithms on the 13-, 123-, and 8500-bus test feeders from state option 2 and reward option 1. The figures report the difference between the reward of the baseline RL algorithms and the reward collected from the test feeders' default control logic. The reward does not have a unit. A difference higher than 0.0 means that the RL agent outperforms the default control logic. The RL algorithms initially need to explore therefore having low reward. However, after about 500 steps of agent-environment interaction, the RL agents are able to outperform the standard control logic. Note that state option 2 does not include nodal power measurements except the substation bus. However, due to the presence of time variables and the temporal patterns of loads, RL can still outperform the default controls.

\begin{figure}[h]
	\centering
	\includegraphics[width=8.8cm]{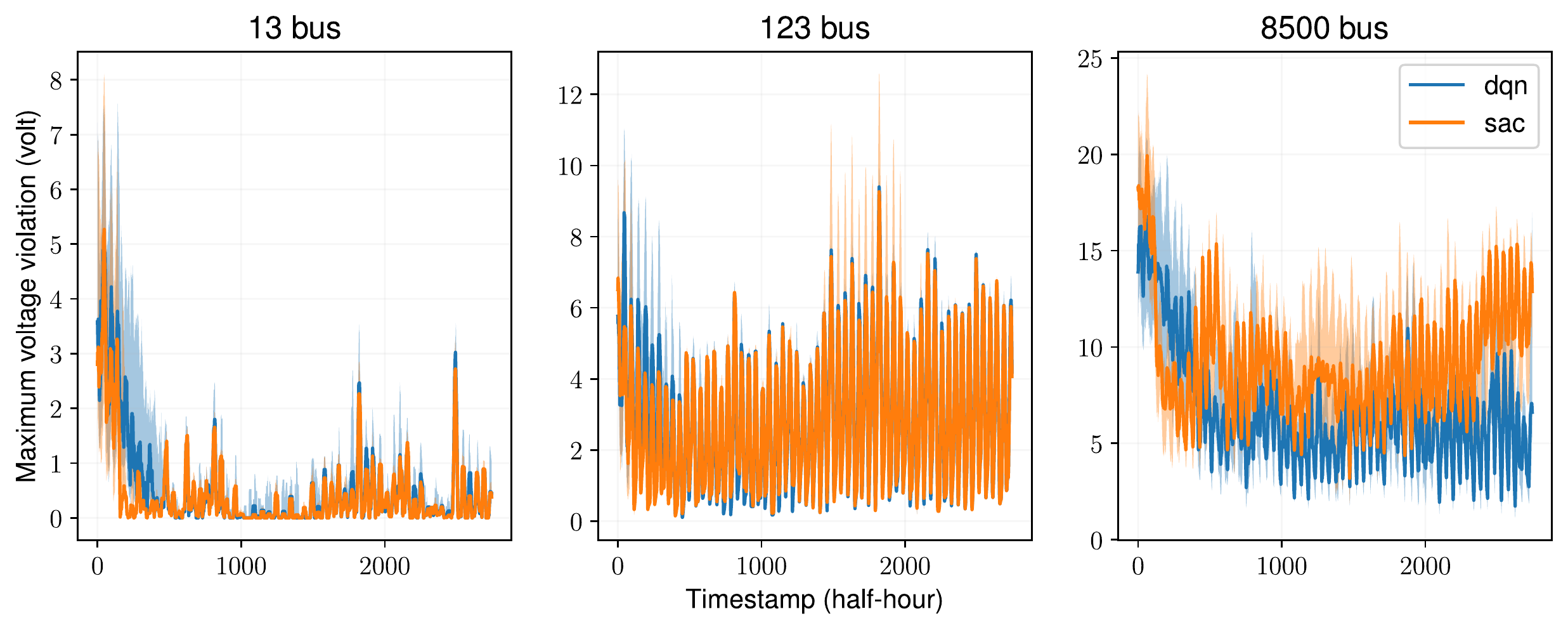}
	\caption{Maximum voltage magnitude violations.}
	\label{fig_voltvio}
\end{figure}

Figure \ref{fig_voltvio} shows the maximum voltage magnitude violations for each timestamp. Although they decrease as training progresses, initially there is a long period of unsafe explorations that produces large voltage magnitude violations. In addition, lowering the network loss and reducing the voltage deviation are two conflicting control tasks, which results in long-term continued voltage violations. In the future, algorithms with safe exploration capabilities should be developed to ensure minimum constraint violations throughout the agent-environment interaction.

\section{Conclusion}
This paper offers an open-source testing environment and dataset for researchers and practitioners to develop and evaluate safe and sample efficient RL-based VVC algorithms. The test feeders and historical operational dataset are created based on the authors' experience and collaboration with the Riverside Public Utility company. We hope the benchmark can serve as a testbed to rigorously evaluate the performance of different RL-based VVC algorithms and accelerate the adoption of data-driven control technology in power distribution systems.
\bibliographystyle{IEEEtran}
\bibliography{RLdataset}

\end{document}